\documentclass[%
 reprint,
superscriptaddress,
 amsmath,amssymb,
 aps,
]{revtex4-1}

\usepackage{graphicx}
\usepackage{dcolumn}
\usepackage{bm}
\usepackage{booktabs}
\usepackage{longtable}
\usepackage{tablefootnote}
\usepackage{threeparttable}

\begin{document}

\preprint{APS/123-QED}

\title{Constraints on key $^{17}$O$(\alpha,\gamma)^{21}$Ne resonances and impact on the weak s-process}

\author{M. Williams}
 \email{mwilliams@triumf.ca}
\affiliation{TRIUMF, 4004 Wesbrook Mall, Vancouver, BC, Canada, V6T 2A3}
\affiliation{Department of Physics, University of York, Heslington, York, YO10 5DD, United Kingdom}

\author{A.M. Laird}
\affiliation{Department of Physics, University of York, Heslington, York, YO10 5DD, United Kingdom}

\author{A. Choplin}
\affiliation{Institut d'Astronomie et d'Astrophysique, Universit\'e Libre de Bruxelles,  CP 226, B-1050 Brussels, Belgium}

\author{P. Adsley}
\affiliation{Cyclotron Institute, Texas A\&M University, College Station, Texas 77843, USA}
\affiliation{Department of Physics \& Astronomy, Texas A\&M University, College Station, Texas 77843, USA}

\author{B. Davids}
\affiliation{TRIUMF, 4004 Wesbrook Mall, Vancouver, BC, Canada, V6T 2A3}
\affiliation{Department of Physics, Simon Fraser University, 8888 University Drive, Burnaby, BC, V5A 1S6, Canada}

\author{U. Greife}
\affiliation{Department of Physics, Colorado School of Mines, Golden, Colorado 80401, USA}

\author{K. Hudson}
\affiliation{Department of Physics, Simon Fraser University, 8888 University Drive, Burnaby, BC, V5A 1S6, Canada}
\affiliation{TRIUMF, 4004 Wesbrook Mall, Vancouver, BC, Canada, V6T 2A3}

\author{D. Hutcheon}
\affiliation{TRIUMF, 4004 Wesbrook Mall, Vancouver, BC, Canada, V6T 2A3}

\author{A. Lennarz}
\affiliation{TRIUMF, 4004 Wesbrook Mall, Vancouver, BC, Canada, V6T 2A3}
\affiliation{Department of Physics and Astronomy, McMaster University, Hamilton, ON, L8S 4L8, Canada}

\author{C. Ruiz}
\affiliation{TRIUMF, 4004 Wesbrook Mall, Vancouver, BC, Canada, V6T 2A3}

\date{\today}

\begin{abstract}

The efficiency of the slow neutron-capture process in massive stars is strongly influenced by neutron-capture reactions on light elements. At low metallicity, $^{16}$O is an important neutron absorber, but the effectiveness of $^{16}$O as a light-element neutron poison is modified by competition between subsequent $^{17}$O$(\alpha,n)^{20}$Ne and $^{17}$O$(\alpha,\gamma)^{21}$Ne reactions. The strengths of key $^{17}$O$(\alpha,\gamma)^{21}$Ne resonances within the Gamow window for core helium burning in massive stars are not well constrained by experiment. This work presents more precise measurements of resonances in the energy range $E_{c.m.} = 612 - 1319$ keV. We extract resonance strengths of $\omega\gamma_{638} = 4.85\pm0.79$ $\mu$eV, $\omega\gamma_{721} = 13.0^{+3.3}_{-2.4}$ $\mu$eV, $\omega\gamma_{814} = 7.72\pm0.55$ meV and $\omega\gamma_{1318} = 136\pm 13$ meV, for resonances at $E_{c.m.} =$ 638, 721, 814 and 1318 keV, respectively. We also report an upper limit for the 612 keV resonance of $\omega\gamma<140$ neV ($95\%$ c.l.), which effectively rules out any significant contribution from this resonance to the reaction rate. From this work, a new $^{17}$O$(\alpha,\gamma)^{21}$Ne thermonuclear reaction rate is calculated and compared to the literature. The effect of present uncertainties in the $^{17}$O$(\alpha,\gamma)^{21}$Ne reaction rate on weak s-process yields are then explored using post-processing calculations based on a rotating $20M_{\odot}$ low-metallicity massive star. The resulting $^{17}$O$(\alpha,\gamma)^{21}$Ne reaction rate is lower with respect to the pre-existing literature and found to enhance weak s-process yields in rotating massive star models.

\end{abstract}

\pacs{Valid PACS appear here}
\maketitle

\section{Introduction}

The weak component of the astrophysical s-process is responsible for producing elements in the A=60--90 mass range \cite{Kappeler2011}. The weak s-process occurs in massive stars mainly during core helium  burning and  to a lesser extent during convective carbon shell burning, with the $^{22}$Ne$(\alpha,n)^{25}$Mg reaction as the main source of free neutrons. During core helium burning, rotation-induced mixing between the helium burning core and hydrogen burning shell results in a primary mechanism for producing $^{14}$N, which can then undergo successive $\alpha$-captures to produce $^{22}$Ne \cite{Pignatari_2008}. Models of rotating massive stars with low metallicity show significant enhancement in s-process yields with respect to non-rotating models \cite{Frischknecht2015,choplin2018,Limongi_2018,Banerjee_2019}. Furthermore, enhanced s-process yields at low metallicity offer an explanation for abundance ratios observed in extremely metal-poor stars, such as those found in the galactic bulge \cite{chiappini2011}. 

Heavy-element yields from the s-process are affected by the role of so-called light element neutron poisons, such as: $^{12}$C, $^{14}$N, $^{16}$O, $^{20}$Ne and $^{25}$Mg \cite{Prantzos1990,Rayet2000}. These elements can undergo neutron-capture reactions, thereby reducing the number of neutrons available for capture by iron-peak seed nuclei. In metal-poor environments, the $^{16}$O$(n,\gamma)^{17}$O reaction makes $^{16}$O a strong neutron poison, due to the high abundance of primary $^{16}$O relative to other neutron poisons. However, the captured neutrons can be subsequently released back into the star via the $^{17}$O$(\alpha,n)^{20}$Ne reaction. Therefore, the competition between $^{17}$O$(\alpha,n)$ and $^{17}$O$(\alpha,\gamma)$ reaction channels is important for determining the role of $^{16}$O as a light-element neutron poison \cite{baraffe1992}. Indeed, it has been shown that predicted yields of s-process elements between Sr and Ba are highly sensitive to changes in the assumed $^{17}$O$(\alpha,\gamma)^{21}$Ne rate \cite{Frischknecht2015,choplin2018}.

The $^{17}$O$(\alpha,\gamma)^{21}$Ne and $^{17}$O$(\alpha,n)^{20}$Ne reactions have been studied directly by Best \textit{et al.} \cite{best2011,best2013}. The $^{17}$O$(\alpha,n)^{20}$Ne reaction was measured by Best \textit{et al.} \cite{best2013} in the center-of-mass energy range of 0.65 to 1.86 MeV, which is just above the Gamow windows for core helium burning of 0.3 to 0.65 MeV in the temperature range between 0.2 and 0.3 GK. Three strong $^{17}$O$(\alpha,\gamma)^{21}$Ne resonances were measured by Best \textit{et al.} \cite{best2011} at $E_{c.m.} =$ 811, 1122, and 1311 keV. The available data on $^{17}$O$(\alpha,\gamma)^{21}$Ne were later extended to lower energies by Taggart \textit{et al.} \cite{taggart2019} down to $E_{c.m.} =$ 633 keV, and also included a scan over the 811 keV resonance. However, the lowest energy data point covered an energy range that contained 3 known $^{21}$Ne states within the target at $E_{x} =$ 7982, 7980 and 7961 keV. Due to low statistics, Taggart \textit{et al.} conclude that it was not possible to determine which of two potential resonances at $E_{c.m.} =$ 613 ($E_{x} =$ 7961 keV) and 634 keV ($E_{x} =$ 7982 keV) dominated the observed yield. The other state at 7980 keV corresponds to a neutron-emitting resonance which, according to the strength estimates of Best \textit{et al.} \cite{best2013}, would not compete with either the 613 or 634 keV resonances in the $(\alpha,\gamma)$ channel. The authors note that if the 613 keV were to dominate over the 634 keV resonance then the reaction rate would be increased by a factor of 2.25 at the relevant temperatures for core helium burning. The authors' choice to adopt a resonance energy of 634 keV was guided by the calculations presented by Best \textit{et al.} \cite{best2013}, which predict the 613 keV resonance to be an order of magnitude weaker than the 634 keV resonance.

The resonance strength values quoted by Taggart \textit{et al.} \cite{taggart2019} also carry large uncertainties of 75\% and 80\% for the 634 and 721 keV resonances, respectively. Estimates for the strengths of presently unmeasured lower energy resonances were presented by Best \textit{et al.} \cite{best2013}, which rely on calculated single-particle widths and order-of-magnitude estimates of $\Gamma_{\gamma}/\Gamma_{n}$ based on systematics. Thermonuclear rate calculations based on these aforementioned estimates and all available data prior to this publication suggest that the strength of a potential resonance at $E_{c.m.} = 305$ keV, which corresponds to a $J^{\pi}=7/2^{+}$ state at $E_{x} = 7648$ keV, remains the dominant source of uncertainty in the rate within the Gamow window for core helium burning. The 305 keV resonance is presently beyond the reach of direct measurements, with an estimated strength of $\omega\gamma=40$ peV. However, if the $\alpha$-width of the $E_{x} = 7648$ keV state is significantly smaller than its estimated value, then resonances at 613, 634 and 811 keV may dominate the high temperature portion of the Gamow window, and so more precise measurements of these resonances are highly desirable.

\section{Experiment Description}

The $^{17}$O$(\alpha,\gamma)^{21}$Ne reaction was studied in inverse kinematics using the DRAGON facility \cite{Hutcheon2003} located in the ISAC-I experimental hall at TRIUMF. DRAGON comprises a windowless differentially pumped gas target surrounded by a $4\pi$ array of 30 BGO scintillators, coupled to an electromagnetic recoil separator. The apparatus used here is similar to that described for a previous study of this reaction performed by Taggart \textit{et al.} \cite{taggart2019}, except the focal plane end-detector for which we instead use a DSSD (Double-sided Silicon Strip Detector) to stop the $^{21}$Ne recoils rather than an ionization-chamber. The downstream Micro Channel Plate (MCP) detector, which forms the stop signal of a local transmission time-of-flight system between two MCPs,  was absent with respect to the previous set-up. However, the time difference between events recorded by the upstream MCP (MCP0) and the DSSD provides a similar local time-of-flight measurement, albeit with poorer timing resolution owing to the DSSD. 

The $^{17}$O$^{3+}$ beam was produced by the TRIUMF Off-Line Ion-Source (OLIS) facility using the multi-charge ion-source \cite{Jayamanna2010}. The average beam intensity delivered for the present work was higher than Taggart \textit{et al.} \cite{taggart2019}, owing to a new RF-buncher cavity installed prior to the radio-frequency quadrupole (RFQ) accelerator (the first acceleration stage at ISAC). The RF-buncher was used to effectively increase the acceptance of the RFQ to allow a greater number of beam-ions per beam packet. The average beam intensity was thus increased to $4.23 \times 10^{12}$ s$^{-1}$ (or 2 $\mu$A) at the experiment, a factor 3.3 increase over the previous study. A total of 6 yield measurements were performed on 5 resonances located at: 1311, 811, 721, 634 and 613 keV, in the center-of-mass frame. Two measurements were performed on the 811 keV resonance at two different gas pressures: 5.6 Torr and 4.2 Torr. The $^{21}$Ne recoils were identified by their energy deposited in the DSSD end-detector, in conjunction with their time-of-flight between the MCP and DSSD. Further selectivity was gained by measuring the time between promptly emitted $\gamma$-rays from the reaction, detected by a $4\pi$ BGO array surrounding the gas target, and heavy-ions detected at the focal plane. For details on the data acquisition system (DAQ) the reader is referred to Ref. \cite{Greg2014}. 

Significantly improved background suppression with respect to the previous study \cite{taggart2019} was gained by tuning the separator to the $3^{+}$ and $2^{+}$ recoil charge states, depending on the targeted resonance, as opposed to the $4^{+}$ recoil charge state selected by Taggart \textit{et al.} \cite{taggart2019}. This is because the $4^{+}$ recoil charge state is closer in mass/charge $(m/q)$ ratio to the $^{17}$O$^{3+}$ unreacted beam, which predominantly emerges from the target in the $3^{+}$ charge state over the studied energy range. 

\section{Data Analysis} \label{sec:data_analysis}

The data were analyzed similarly to the previous study of this reaction \cite{taggart2019}. The measured strength of a compound nuclear resonance is given by:

\begin{equation} \label{eqn:wg1}
\begin{split}
    \omega\gamma = & \frac{2\pi\epsilon(E_{r})Y}{\lambda_{r}^{2}} \; \times \\ & \Bigg [\mathrm{arctan}\Bigg(\frac{E_{0}-E_{r}}{\Gamma/2} \Bigg ) - \mathrm{arctan}\Bigg(\frac{E_{0}-E_{r}-\Delta E}{\Gamma/2} \Bigg ) \Bigg ]^{-1}
\end{split}
\end{equation} 

where $\epsilon(E_{r})$ is the beam stopping power at the resonance energy $E_{r}$, $Y$ is the reaction yield per incident beam ion, $\lambda_{r}$ is the de Broglie wavelength of the system at $E_{r}$, $E_{0}$ is the incident beam energy, $\Delta E$ is the energy loss across the target, and $\Gamma$ is the total width of the resonance. All energies are in the center of mass frame. In cases where the resonance width is narrow compared to the energy loss over the target $(\Gamma << \Delta E)$, the thick target approximation may be applied, reducing Equation \ref{eqn:wg1} to:

\begin{equation} \label{eqn:wg2}
    \omega\gamma = \frac{2\epsilon(E_{r})Y_{\infty}}{\lambda_{r}^{2}}.
\end{equation}
 
The beam stopping power was obtained directly by measuring the beam energy before and after filling the target with helium gas. These energy measurements were performed by tuning the beam through DRAGON's first magnetic dipole and then calculating the energy using Equation 2 in Ref. \cite{Hutcheon2012}. The uncertainty in each measurement of the incoming and outgoing beam energy was 0.15\%, determined by the MD1 magnetic field constant \cite{Hutcheon2012}. The relative uncertainty in the energy loss was in the range of 4 to 7\%, depending on how much energy was lost through the target. The density of the target gas was calculated via the ideal gas law from the measured pressure and temperature of the target gas in addition to the known effective length of the gas target: $L_{\mathrm{eff}} =12.3\pm 0.4$ cm \cite{Hutcheon2003}. The error in the effective length of the gas target dominates the uncertainty in the number density. Both the target number density and energy loss uncertainties factor into the calculation of beam stopping power; therefore, the total uncertainty in the stopping power ranged between 6 and 8\%. The reaction yield per incident beam ion is given by:

\begin{equation}
    Y = \frac{N_{r}}{N_{b}\varepsilon},
\end{equation}

where $N_{b}$ is the total number of beam ions on target and $N_{r}$ is the number of detected reaction events with detection efficiency $\varepsilon$. The overall detection efficiency for recoils at the end-detector, without requiring a coincident $\gamma$-ray, is termed the \textit{singles} detection efficiency and is given by:

\begin{equation}
    \varepsilon_{s} = f_{q} \cdot \eta_{r} \cdot \eta_{MCP} \cdot \varepsilon_{MCP} \cdot \varepsilon_{DSSD} \cdot \lambda_{s},
\end{equation}

where $f_{q}$ is the recoil charge state fraction in the tuned recoil charge state $q$, $\eta_{r}$ is the recoil transmission efficiency through DRAGON, $\eta_{MCP}$ is the transmission efficiency through the MCP, $\varepsilon_{MCP}$ is the detection efficiency of the MCP, $\varepsilon_{DSSD}$ is the detection efficiency of the DSSD, and $\lambda_{s}$ is the singles DAQ live-time fraction. Requiring coincident detection of $\gamma$-rays from the reaction means that the BGO detection efficiency, $\varepsilon_{BGO}$, must be included and the DAQ live-time fraction is now the \textit{coincidence} live-time fraction, $\lambda_{c}$. The coincidence efficiency, $\varepsilon_{c}$ is then calculated as:

\begin{equation}
    \varepsilon_{c} = f_{q} \cdot \eta_{r} \cdot \eta_{MCP} \cdot \varepsilon_{MCP} \cdot \varepsilon_{DSSD} \cdot \varepsilon_{BGO} \cdot \lambda_{c}.
\end{equation}

Both the recoil transmission efficiency and the BGO efficiency were determined using a \texttt{GEANT3} simulation of the entire facility \cite{GIGLIOTTI2003}. The mean recoil cone angles for the resonances studied here do not exceed approximately 15 mrad, which is less than the 21 mrad maximum angular acceptance of DRAGON. However, this is a substantial enough cone angle for some losses to occur, for instance, the 814 keV resonance has a calculated recoil transmission of $86 \pm 2\%$ and the 612 keV resonance has recoil transmission of $75 \pm 2\%$. The accuracy of transmission efficiencies obtained using the \texttt{GEANT3} simulation has recently been verified for much larger recoil cone angles, for instance with the $^{6}$Li$(\alpha,\gamma)^{10}$B reaction, which has a cone angle of 32 mrad \cite{psaltis2021}. The systematic uncertainty associated with the simulated detection efficiency of the BGO array is 10.3\% \cite{GIGLIOTTI2003} and only applies to results obtained from coincidence events.

The total number of incident beam ions on target ($N_{b}$) was determined by hourly Faraday cup measurements. Fluctuations in beam current within runs were accounted for by monitoring the scattering rate of recoiling $^{4}$He ions from the target gas, detected by two silicon surface barrier detectors installed in the target chamber. A detailed description of the beam normalisation procedure is provided in Ref. \cite{Williams2020}. The uncertainty in the total beam on target was 1\% for the 814 and 1318 keV resonances, 5\% for the 634 keV resonance and 7\% for the 720 and 612 keV resonances. The uncertainty in beam transmission through the target, as determined by Faraday cups before and after the target (see Ref. \cite{Williams2020} for details on this procedure), dominated the uncertainty in $N_{b}$. The relative uncertainty in the measured beam transmission can vary depending on the stability of the beam current while cup readings are in progress. The MCP transmission efficiency was determined by tuning attenuated beam through DRAGON and recording the DSSD rate with the MCP inserted and retracted from the beamline. The MCP detection efficiency was also determined using attenuated beam runs at each beam energy. The MCP transmission and detection efficiencies were measured as $88 \pm 2 \%$ and $90.1 \pm 3.6\%$, respectively. The $96.15 \pm 0.53$\% geometric efficiency of the DSSD was reported in Ref \cite{Wrede-NIMB204-2003}.

We note here that the previous measurement by Taggart $\textit{et al.}$ \cite{taggart2019} used incorrect recoil charge state fractions adopted from Ref. \cite{taggart2012}. By inspection of Table 5.2 in Ref. \cite{taggart2012}, it appears that the speed of the $^{21}$Ne beam used to measure the charge state distributions was confused with the corresponding value for the $^{17}$O beam during the $^{17}$O$(\alpha,\gamma)^{21}$Ne experiment. This led to an incorrect conversion to center-of-mass energy, and so the wrong charge state fractions were applied to the actual center-of-mass energies considered in the experiment. For example, the $4^{+}$ recoil charge state fraction utilized by Taggart et al. for the 814 keV resonance was $39\%$, which would indeed be the correct value if the recoil speed were 251 keV/$u$ (equivalent to the $^{17}$O beam). However, the recoil speed is in fact around 165 keV/$u$ for this resonance, where the $4^{+}$ recoil charge state fraction drops to $18\%$. Therefore, the strength value of $5.4 \pm 0.8$ meV reported in Taggart \textit{et al.} is increased to $11.7 \pm 1.7$ meV. It is important to note that the underlying charge state distribution measurements were nonetheless correct, and were therefore used for the present work. The uncertainty in the recoil charge state fractions ranged between 5.2 and 8.0\%.

\begin{figure*}[httt]
\minipage{0.33\textwidth}
  \includegraphics[width=1.\textwidth]{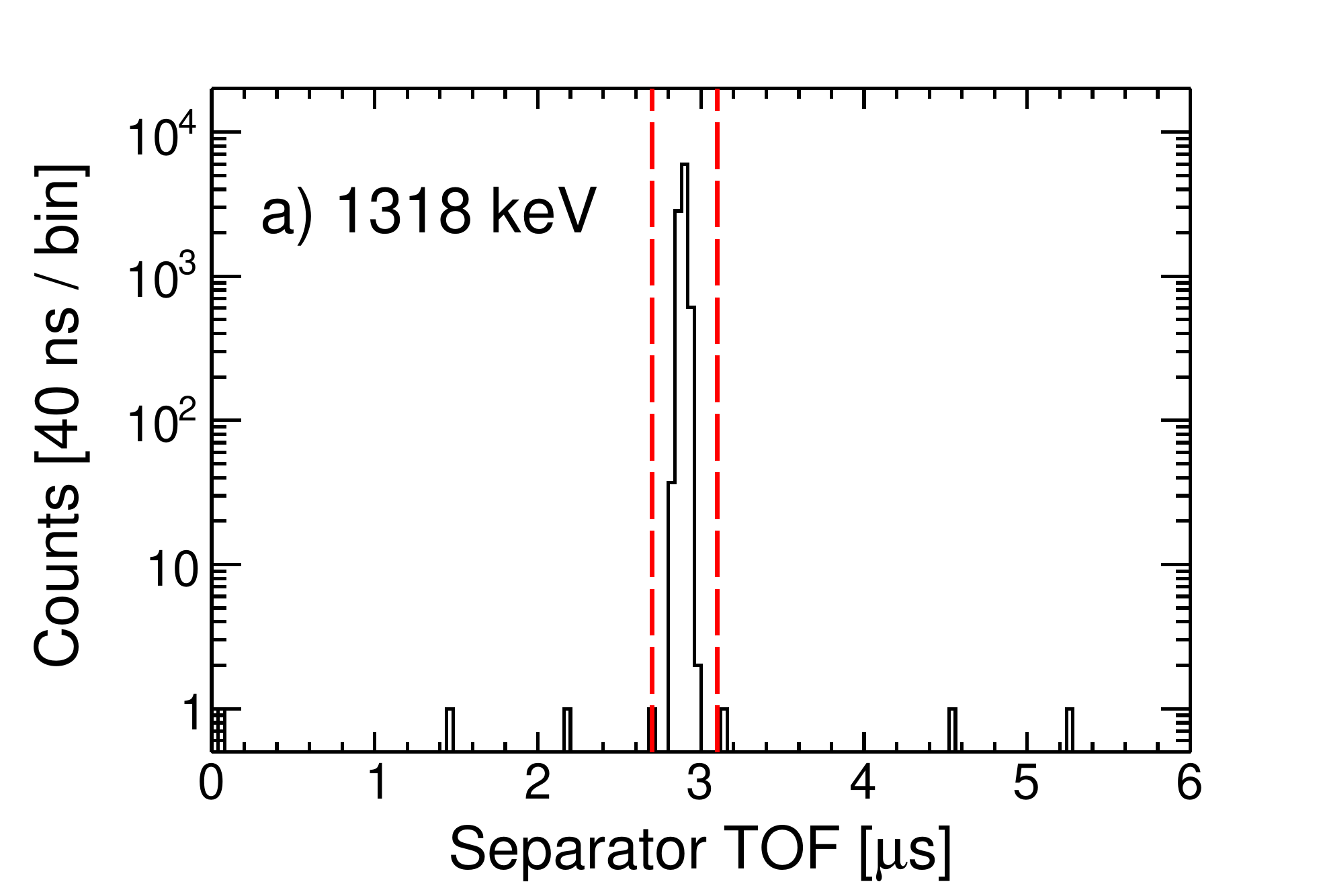}
\endminipage\hfill
\minipage{0.33\textwidth}
  \includegraphics[width=1.\textwidth]{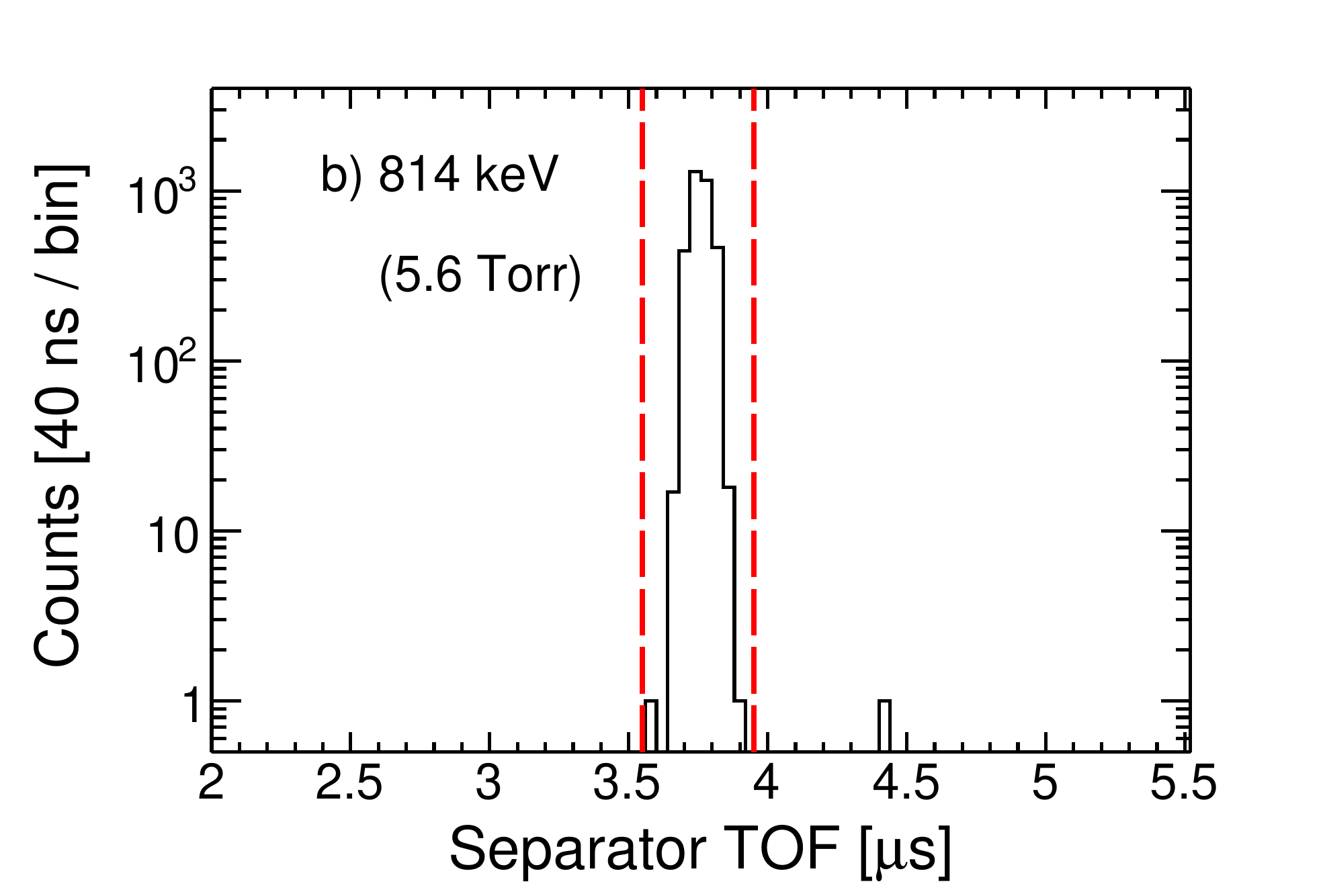}
\endminipage\hfill
\minipage{0.33\textwidth}
  \includegraphics[width=1.\textwidth]{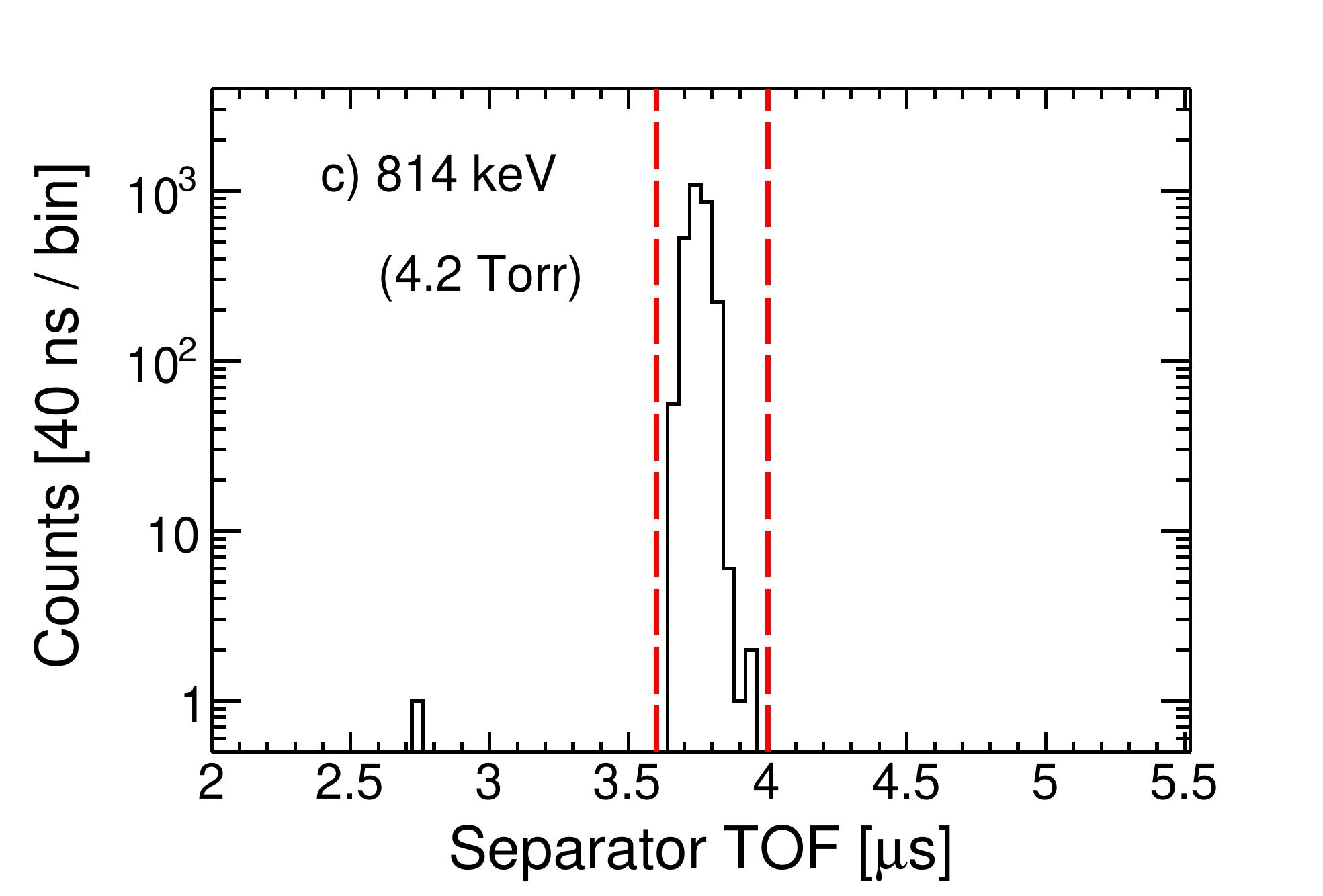}
\endminipage \\
\minipage{0.33\textwidth}
  \includegraphics[width=1.\textwidth]{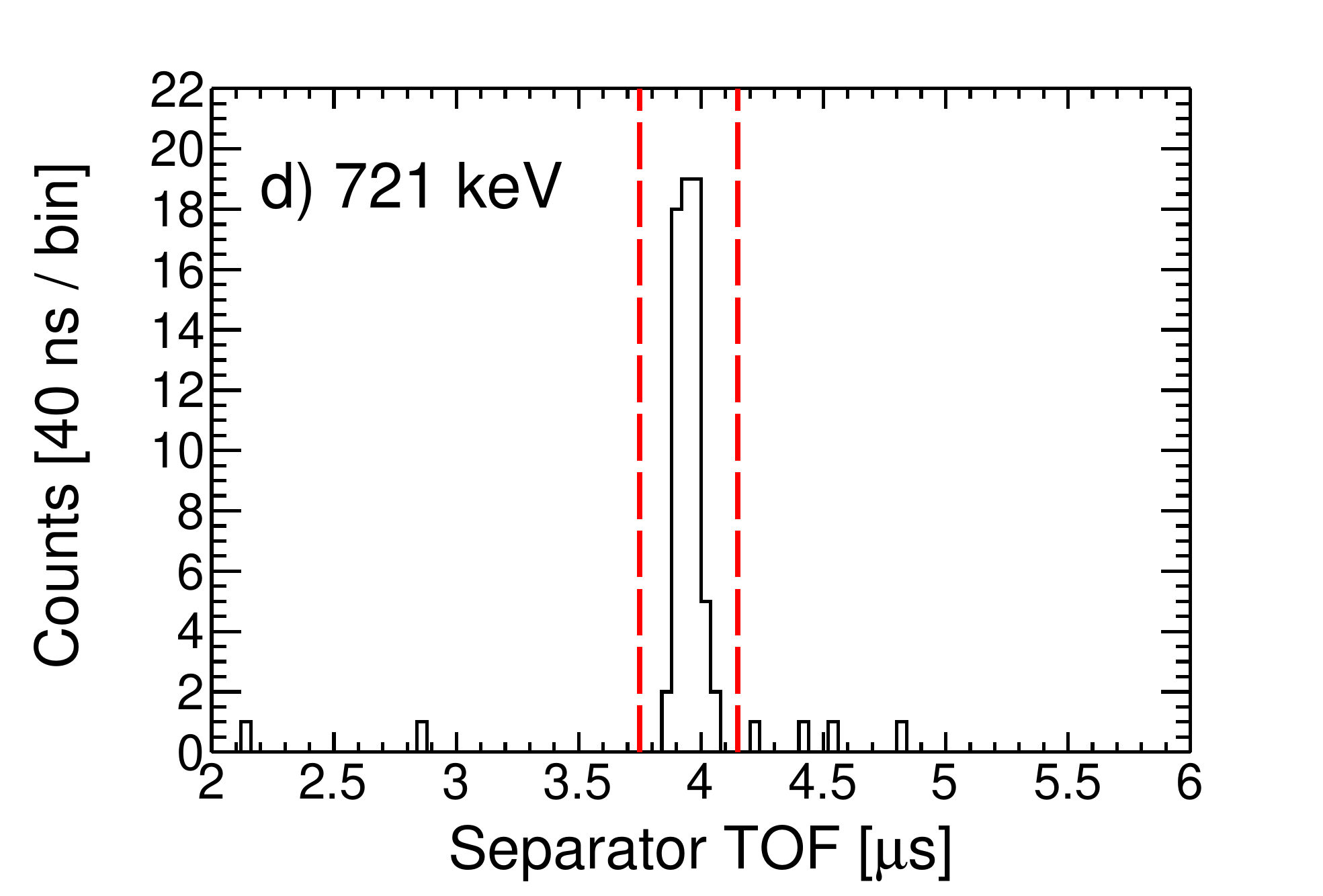}
\endminipage\hfill
\minipage{0.33\textwidth}
  \includegraphics[width=1.\textwidth]{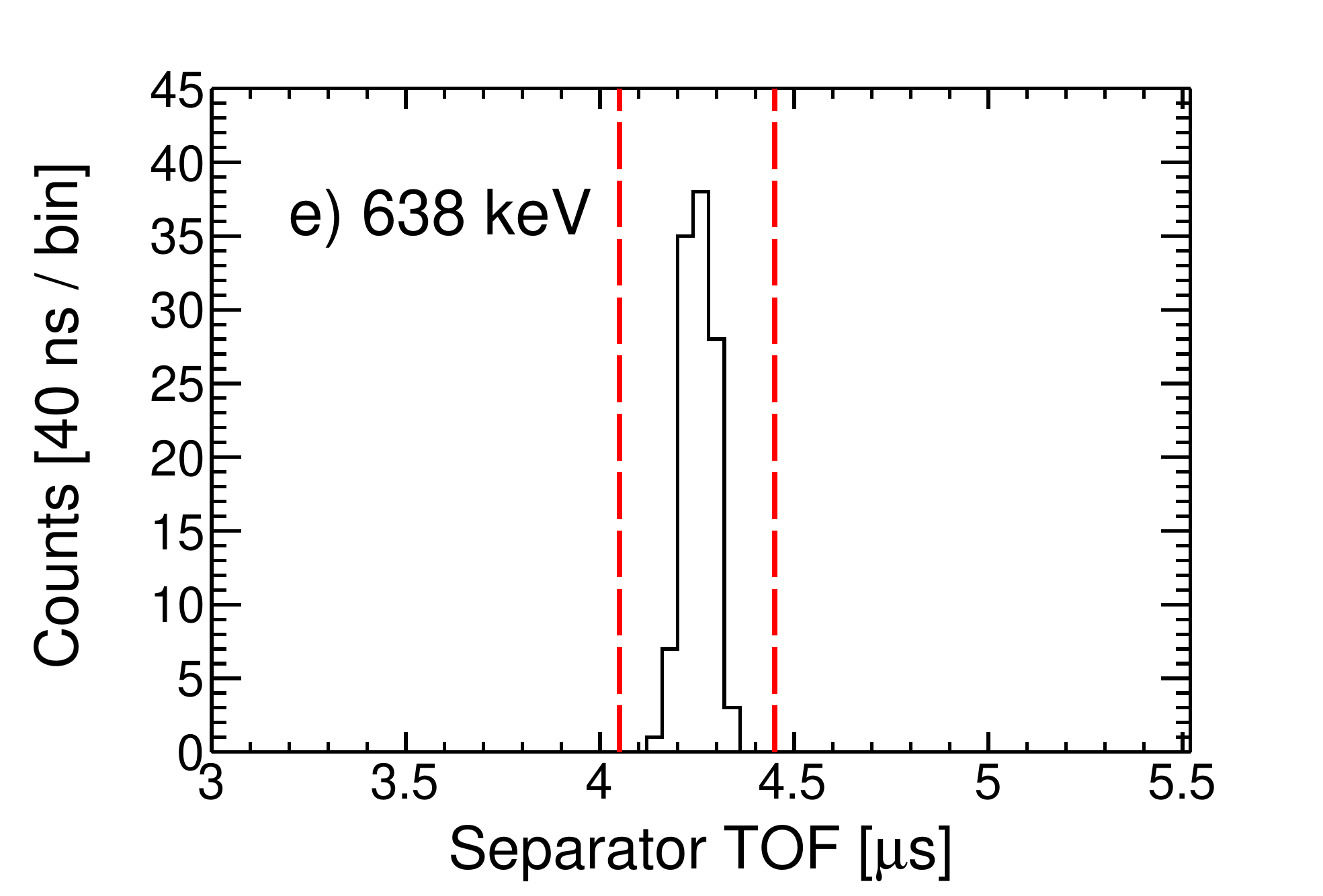}
\endminipage\hfill
\minipage{0.33\textwidth}
  \includegraphics[width=1.\textwidth]{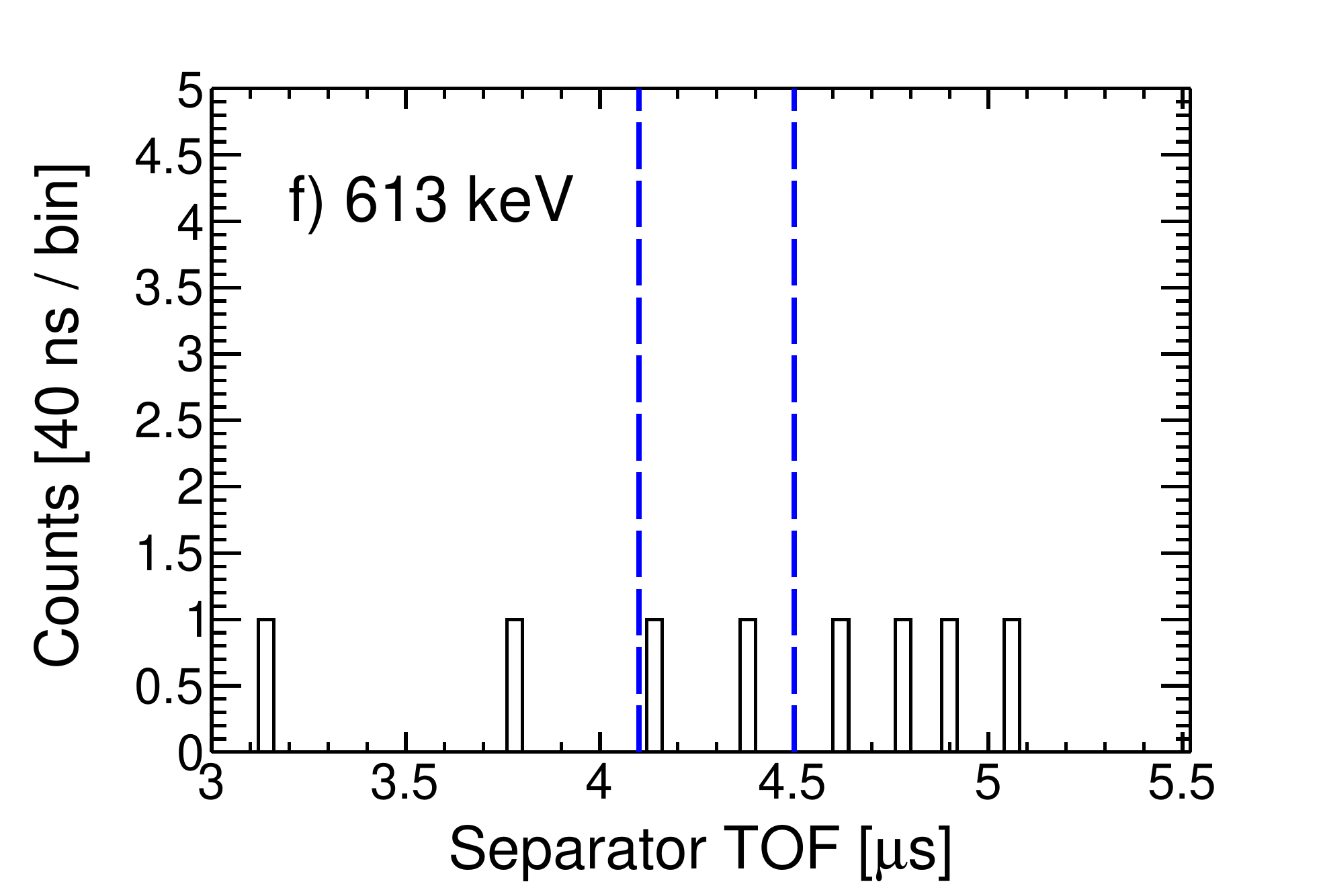}
\endminipage
\caption{Separator time-of-flight (TOF) spectra for resonant yield measurements at $E_{c.m.} =$ 1318 keV (a), 814 keV with 5.6 Torr (b) and 4.2 Torr (c) gas pressure, 721 keV (d), 634 keV (e) and 613 keV (f). The vertical red dashed lines bound the signal region; no signal was evident for the 613 keV yield measurement. The blue vertical dashed lines on the 613 keV plot indicate the predicted signal region based on extrapolation from the higher-lying resonances; only two counts appear within this region.}
\label{fig:septof}
\end{figure*}

\begin{figure}[ht!]
 \centering
 \includegraphics[width=0.45\textwidth]{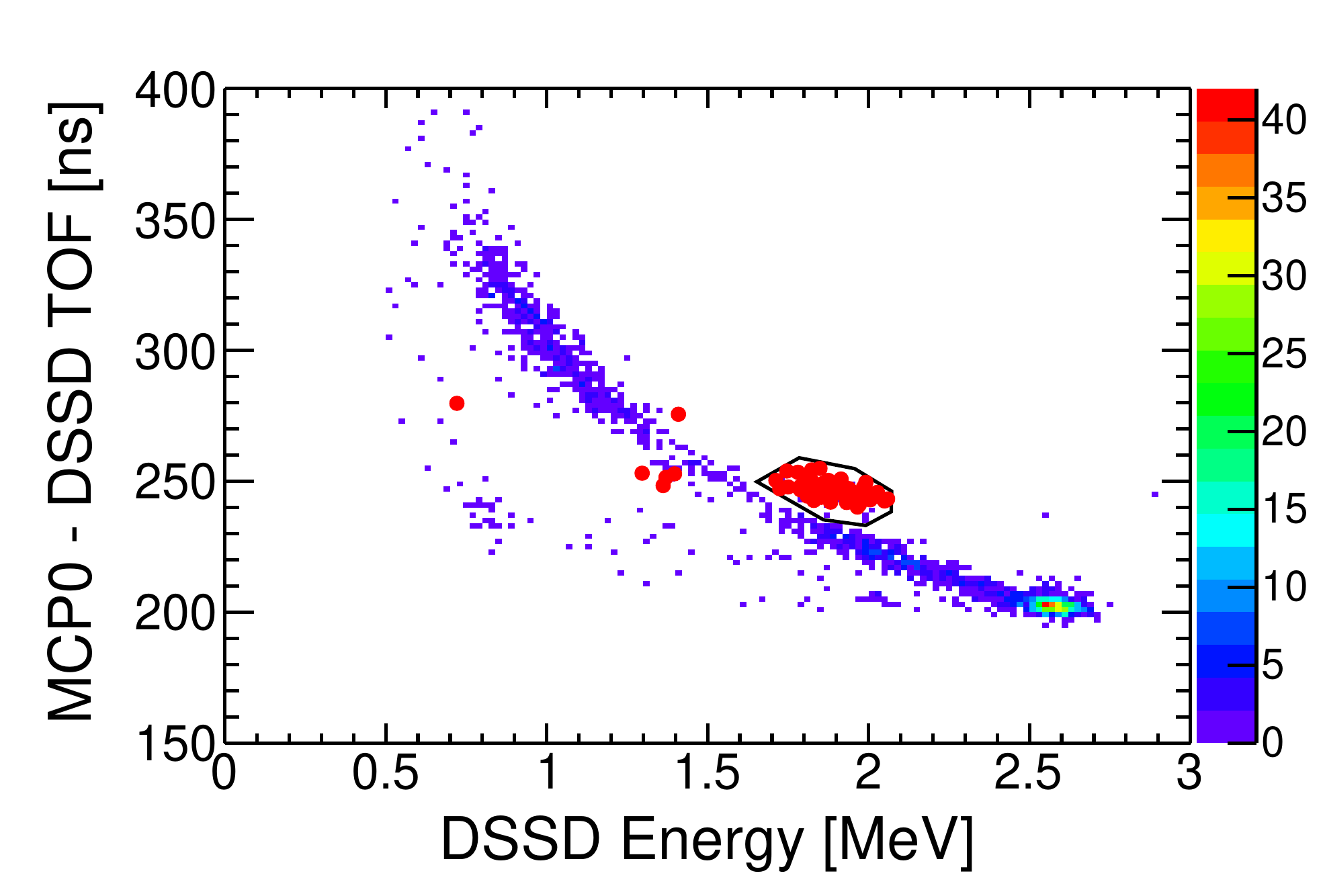}
\caption{MCP-DSSD TOF vs DSSD Energy for the $E_{c.m.} = 634$ keV resonance. The colour plot shows singles data, whereas the red circles indicate coincidence events falling with the separator TOF window. The black lined polygon is a graphical cut around the $^{21}$Ne recoils.} \label{fig:Eres634keV_pid}
 \end{figure}

\begin{figure}[ht!]
 \centering
 \includegraphics[width=0.45\textwidth]{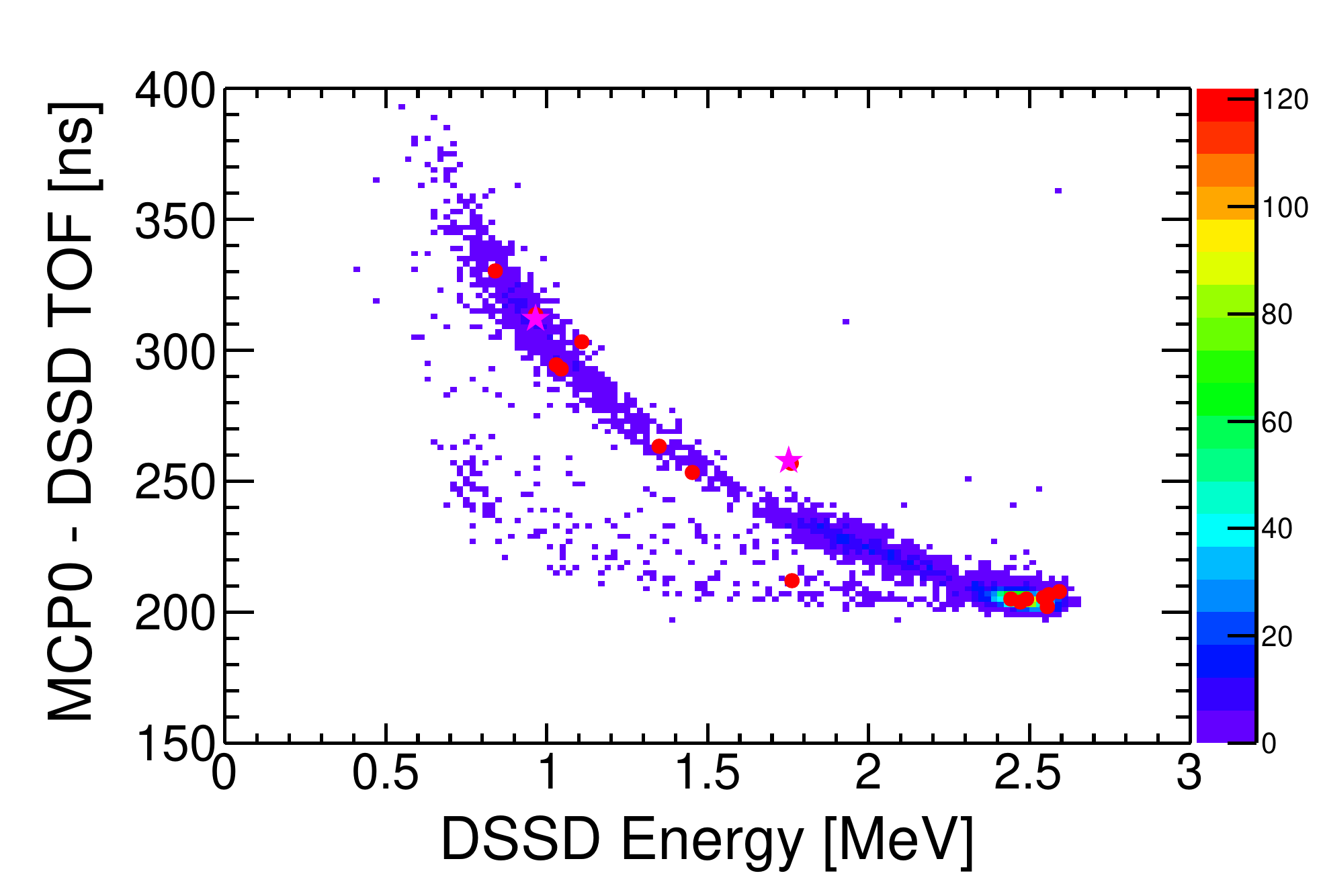}
\caption{MCP-DSSD TOF vs DSSD Energy for the $E_{c.m.} = 613$ keV resonance. The colour plot shows singles data, whereas the red circles indicate coincidence events. The pink stars show the two events falling within the expected separator-TOF signal region} \label{fig:Eres613keV_pid}
 \end{figure}

Figure \ref{fig:septof} shows the separator time-of-flight (TOF): the time difference between detected $\gamma$-rays and recoils, for each yield measurement. The characteristic peak arises due to correlated recoil-$\gamma$ events from the $^{17}$O$(\alpha,\gamma)^{21}$Ne reaction. No peak is observed for the lowest-energy yield measurement. $^{21}$Ne recoils can be unambiguously identified without a coincident $\gamma$-ray by plotting the DSSD-MCP time-of-flight vs DSSD energy, as demonstrated by Figure \ref{fig:Eres634keV_pid} for the 634 keV yield measurement. The red markers indicate coincidence events, which help to identify the region of interest. Figure \ref{fig:Eres613keV_pid} shows the same plot for the yield measurement at $E_{c.m.}=613$ keV. The pink stars indicate the two coincidence events occurring within the extrapolated separator-TOF signal region, which is indicated in panel f) of Figure \ref{fig:septof} as the region bounded by the two vertical blue dashed lines. One of these events appears to exhibit very similar kinematic properties, with respect to scattered beam background, as the $^{21}$Ne recoils observed for the 634 keV resonance. Therefore, with only one prospective count, we prefer to present only an upper limit on the observed yield. Similar plots were used in the analysis of the 814 keV and 721 keV yield measurements. The MCP was not inserted for the 1311 keV yield measurement, since the beam suppression was sufficient at this energy to not require MCP signals; this is supported by the excellent agreement between the singles resonance strength shown in Table \ref{tab:results} and the strength value of $144 \pm 17$ meV we extract from the coincidence data.

In addition to resonance strengths, resonance energies were determined using the BGO hit pattern method, details of which can be found in Ref. \cite{Hutcheon2012}. The hit pattern in the BGO array was used to determine the location of the resonance along the gas target which, combined with the measured incoming and outgoing beam energies, allowed us to calculate resonance energies to typically within a few keV in the centre of mass frame.

\section{Results}

The resonance strengths and energies from this work are listed in Table \ref{tab:results}, along with values from the literature. There is some discrepancy in the resonance energy for the highest energy resonance studied in this work. The resonance energy of $1311 \pm 2$ keV found by Best \textit{et al.} \cite{best2011} is significantly lower than the value adopted by Firestone \cite{FIRESTONE2015}, disagreeing by more than $2\sigma$. The value found in the present work is even higher than the Firestone value, but within $1\sigma$ agreement. From the singles data, we report a resonance strength of $136 \pm 13$ meV, which is in excellent agreement with the value found by Best \textit{et al.}.

\begin{table*}[ht!]
    \centering
    \begin{threeparttable}
    \caption{Resonance energies and strengths for the $^{17}$O$(\alpha,\gamma)^{21}$Ne reaction. Strengths from the present work are shown from both singles and coincidence data, except for the upper limit for the 613 keV resonance, which is obtained only from coincidence data, and the 720 keV resonance, which is obtained only from singles. We adopt the strength values from singles data, except the 814 keV and 1318 keV resonances, for which we adopt a weighted average between the singles values in the present work and those reported by Best \textit{et al.} \cite{best2011}. Two sets of strength values are shown for the 814 keV resonance, reflecting the two measurements at different gas target pressures, the first row  was taken with 5.6 Torr, the second with 4.2 Torr. The resonance energies are all adopted from the present work, except for the lowest energy resonance, which we adopt from Firestone \cite{FIRESTONE2015}.}
    \begin{tabular}{l c r c l c c r}
     \toprule[1pt]\midrule[0.3pt]
    \multicolumn{3}{c}{$E_{r}$ (keV)} & \phantom{abc} & \multicolumn{4}{c}{$\omega\gamma$} \\ \cmidrule[0.3pt]{1-3} \cmidrule[0.3pt]{5-8} \noalign{\smallskip}
     Firestone \cite{FIRESTONE2015} & Best \textit{et al.} \cite{best2011} &  This Work  &        & Best \textit{et al.} \cite{best2011,best2013} & \multicolumn{2}{c}{This Work} & Adopted \\ \cmidrule[0.3pt]{6-7} \noalign{\smallskip}
     & & & & & Singles & Coincidences & \\
     \midrule[0.3pt] \noalign{\smallskip}
    $613.0\pm1.3$  & &  &  & $^{\mathrm{a}}$ $0.24$ $\mu$eV & & $^{\mathrm{b}}<0.14$ $\mu$eV &  $^{\mathrm{b}}<0.14$ $\mu$eV \\
    $634.2 \pm 0.6$ & & $637.7\pm1.6$   &         & $^{\mathrm{a}}$ $8.9$ $\mu$eV & $4.85 \pm 0.79$ $\mu$eV & $4.98 \pm 0.97$ $\mu$eV & $4.85 \pm 0.79$ $\mu$eV \\
    $721 \pm 2$  & & $720.5\pm2.2$   & &  & $13.0^{+3.3}_{-2.4}$ $\mu$eV & & $13.0^{+3.3}_{-2.4}$ $\mu$eV \\ 
    $807.1 \pm 1.0$ & $811 \pm 2$ & $813.7\pm 1.7$ & & $7.6 \pm 0.9$ meV & $7.51 \pm 0.72$ meV & $8.5 \pm 1.1$ meV & $7.69 \pm 0.47$ meV \\
    & & & & & $8.00 \pm 0.84$ meV & $9.4 \pm 1.1$ meV &\\
    $1316 \pm 1$ & $1311\pm2$ & $1318\pm1.9$ & &  $136 \pm 17$ meV & $136 \pm 13$ meV & $144 \pm 17$ meV & $136 \pm 10$ meV \\
	\midrule[0.3pt]\bottomrule[1pt]
    \end{tabular}
    \begin{tablenotes}
        \item[a] Estimated $\omega\gamma$ values from Best \textit{et al.} (2013) \cite{best2013} based on calculated single-particle widths, order-of-magnitude estimates for $\Gamma_{\gamma}/\Gamma_{n}$ based on systematics, and an assumed spectroscopic factor of $S_{\alpha} = 0.01$. No uncertainties in these estimates were specified.
        \item[b] Upper limit evaluated at the 95\% confidence level.
    \end{tablenotes}
    \label{tab:results}
    \end{threeparttable}
\end{table*}

The resonance at approximately 811 keV is particularly important for the $^{17}$O$(\alpha,\gamma)^{21}$Ne reaction at temperatures relevant for convective carbon shell burning. Again, there is some disagreement with respect to the resonance energy. The present result agrees with Best \textit{et al.} \cite{best2011} at approximately the $1\sigma$ level, however, this resonance is associated in Firestone with the $E_{x}=8155$ keV state \cite{FIRESTONE2015}, corresponding to a resonance energy of 807 keV. The resonance strength found here also agrees well with the value reported by Best \textit{et al.} \cite{best2011}, and so we adopt a weighted average between Best \textit{et al.} and the singles result from the present work. There is strong disagreement between the strength value reported by Taggart \textit{et al.} \cite{taggart2019} and both the present work and Best \textit{et al.} \cite{best2011}, regardless of whether one adjusts for the incorrect charge state fraction used by Taggart \textit{et al.}, which is discussed in Section \ref{sec:data_analysis}. The strength value presented by Taggart \textit{et al.} for the 811 keV resonance was obtained only using coincidence data, since there was too much background to identify recoils without coincident $\gamma$-rays. In that work, the number of coincidences was evaluated with the requirement that at least one $\gamma$-ray be detected with an energy above 2 MeV. This is above the strongly populated 1395 keV $\gamma$-ray from the second-excited state in $^{21}$Ne, which collects much of the strength of higher levels. The coincidence strengths for the present work, evaluated with a $\gamma$-ray energy threshold of 1.2 MeV, are in reasonable agreement with the singles values given in Table \ref{tab:results}. However, imposing a 2 MeV threshold and adopting the same coincidence efficiency as Taggart \textit{et al.} gives a result of $\omega\gamma = 11.5 \pm 1.3$ meV, which is in agreement the result of Taggart \textit{et al.} of $11.7 \pm 1.7$ meV after correcting for the incorrect recoil charge state fractions, but strongly disagrees with the singles result from this work. This strongly suggests that the adopted primary branching ratios are inaccurate, which impacts the calculated efficiency if one imposes a cut that excludes strong secondary transitions. Therefore, due to this issue and the incorrect charge state fractions, the present work supersedes the work of Taggart \textit{et al.}

No coincidence result is reported here for the 720 keV resonance due to unknown $\gamma$-ray branching ratios and too few counts (50 observed coincidence events) to obtain a good estimate of branching ratios by fitting the BGO spectra with simulated decay schemes, as was performed in Ref. \cite{Williams2021} for the $^{19}$F$(p,\gamma)^{20}$Ne reaction. Our strength value is found to be in agreement with the value of  $8.7^{+7.0}_{-3.7}$ $\mu$eV reported by Taggart \textit{et al.}, though the error quoted by Taggart \textit{et al.} is very large. The systematic error bar is considerably larger for this resonance due mainly to the influence of unknown branching ratios on the recoil transmission efficiency. For a $100\%$ branch to the ground-state the transmission could be as low as $40\%$, but could reach $>95\%$ for a cascade of $\gamma$-rays. The central value was estimated to be $82\%$ based on a linear extrapolation of the calculated transmission efficiencies for each resonance as a function of center of mass energy. The upper error was adopted as $+6\%$ such that the transmission would be effectively $100\%$ at $+3\sigma$ deviation. An arbitrary estimate of $-15\%$ was chosen for the lower error limit, which we deem as highly conservative. In any case, this resonance does not contribute appreciably to the overall reaction rate. 

This is the first time that the resonance at 613 keV has been targeted by a direct measurement. One potential count was observed within the expected separator TOF and particle-ID region of interest from which we extract the upper limit given in Table \ref{tab:results} at the $95\%$ confidence level using the method presented in Ref \cite{rolke2001}, and implemented using the \texttt{TRolke} class of the cern ROOT analysis package \cite{brun1997root}. Here we assume a Poisson background model. This result allows us to conclude with certainty that the yield observed for the lowest energy measurement performed by Taggart \textit{et al.} \cite{taggart2019} was due almost entirely to the 634 keV resonance, as was assumed by the authors. The strength reported here for the 634 keV resonance is found to be in agreement with the value presented by Taggart \textit{et al.}, but with significantly reduced uncertainty. The resonance energy is found to be 3.5 keV higher than the value listed in Ref. \cite{FIRESTONE2015}, and so will be referred to as the 638 keV resonance hereafter. The thermonuclear rate presented adopts the resonance energies from the present work. 

\section{Thermonuclear Reaction Rate} \label{sec:reaction_rate}

A new $^{17}$O$(\alpha,\gamma)^{21}$Ne thermonuclear reaction rate was calculated utilizing the resonance energies taken from this work and the adopted resonance strengths listed in Table \ref{tab:results}. The resonance strength of the 1122 keV resonance is adopted from Best \textit{et al.} \cite{best2011}. Parameters for unmeasured resonances were taken from Table 2 of Best \textit{et al.} \cite{best2013}. We note that there are some inconsistencies in Ref. \cite{best2013} for the $\omega\gamma$ values (e.g. the $E_{r} = 305$ keV resonance), whereby the listed resonance strengths contradict the estimated $\Gamma_{\gamma}/\Gamma_{n}$ ratios. For instance, assuming the calculated single-particle $\alpha$-widths given in Table 2 of Best \textit{et al.} \cite{best2013}, which correspond to the lowest possible angular momentum transfer, and an assumed partial width ratio of $\Gamma_{\gamma}/\Gamma_{n} =2$, the 305 keV resonance strength would be $\omega\gamma = 20$ peV, not the $\omega\gamma = 40$ peV presented. We understand this to be the result of a simple numerical error in the table that was propagated to the reaction rate calculation, but that the single-particle widths are indeed correct \cite{Best_PC}.

The present rate, tabulated in Table \ref{table:reactionrate} located in Appendix \ref{sec:reactionrate}, was computed using the \texttt{RatesMC} Monte-Carlo code \cite{iliadis2015}. The contributions of unmeasured resonances were calculated as described in Ref. \cite{longland2012}, with the upper limits in $\Gamma_{\alpha}$ set to the single-particle limits given by Best \textit{et al.} \cite{best2013}, including for the 305 keV resonance. Therefore, our rate corrects for the resonance strength calculation error in Table 2 of Best \textit{et al.} However, in the case of the 612 keV resonance, for which we present an experimental upper limit, the $\Gamma_{\alpha}$ is reduced from the single-particle limit of $12$ $\mu$eV to an upper limit of 105 neV (95\% c.l.) implied by our measurement of $\omega\gamma$, assuming a width ratio of $\Gamma_{\gamma}/\Gamma_{n} =2$ as estimated by Best \textit{et al.} \cite{best2013}.

\begin{figure}[ht!]
 \centering
 \includegraphics[width=0.45\textwidth]{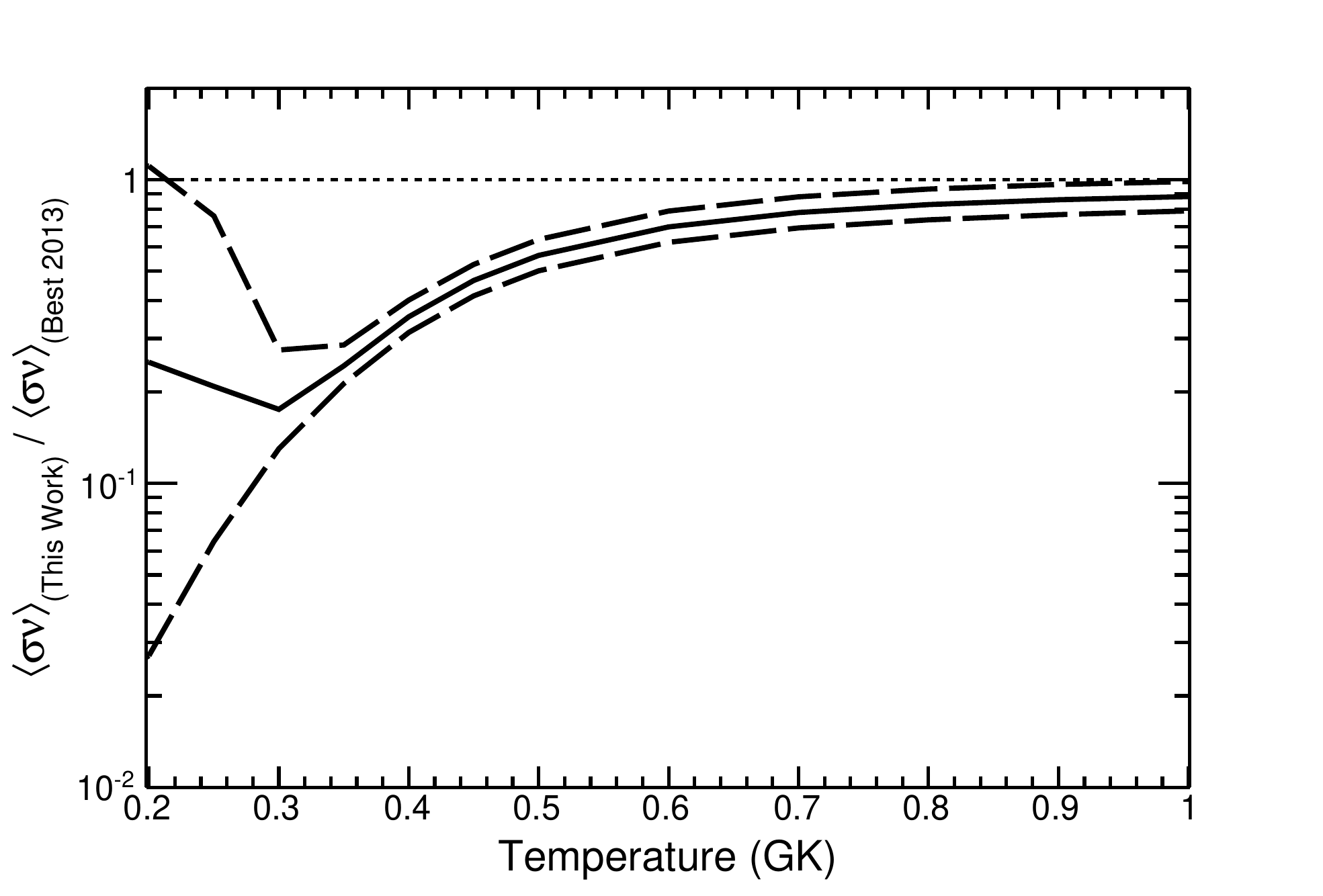}
\caption{Recommended median (black solid line), upper and lower rates (black dashed lines) expressed as a ratio to the tabulated recommended rate in Best \textit{et al.} \cite{best2013}.}
\label{fig:rate_compare}
\end{figure}

\begin{figure}[ht!]
 \centering
 \includegraphics[width=0.45\textwidth]{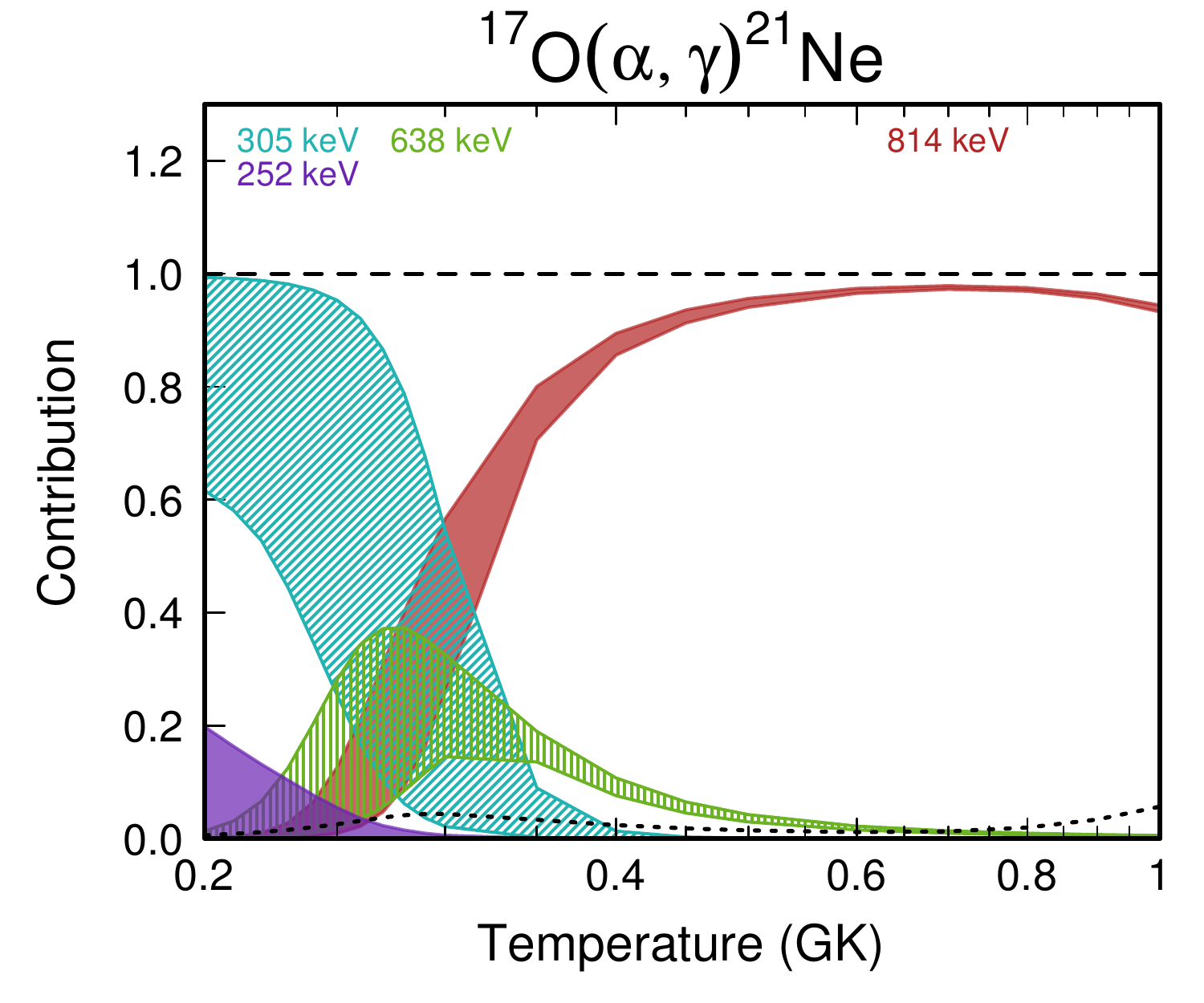}
\caption{Contribution of resonances to the total reaction rate. Only resonances that contribute at least $10\%$ to the total rate within the displayed range are indicated with coloured bands; the sum of all other contributing resonances are represented by the black dotted line. The upper limit we place on the strength of the 613 keV resonance effectively excludes any significant contribution from this resonance.} \label{fig:rate_contribution}
\end{figure}

Figure \ref{fig:rate_compare} displays the present median rate and uncertainties expressed as a ratio of the tabulated rate recommended by Best \textit{et al.} \cite{best2013}. Uncertainties are presented as the 16$^{\mathrm{th}}$ and 84$^{\mathrm{th}}$ percentile of the cumulative reaction rate distribution for the low and high rate, respectively. The present rate is lower at all temperatures, with a factor of 4 to 5 reduction in the temperature region of interest between 0.2 and 0.3 charge
GK for core helium burning.  While our results constrain the rate at the upper temperature range of the Gamow window, the dominant source of uncertainty at these temperatures remains the 305 keV resonance, which is presently inaccessible to direct measurements. The contribution of each resonance is shown by Figure \ref{fig:rate_contribution}, which shows that the upper limit extracted for the 613 keV resonance effectively removes any significant contribution from this resonance. The 638 keV resonance, for which we report a reduction in uncertainty from 80\% to 25\%, may contribute up-to approximately 40\% within the relevant temperature range. This contribution may be greater if the $\alpha$-width of the 305 keV resonance were measured to be significantly below its estimated value.

\section{Astrophysical Impact}

The impact of present uncertainties in the $^{17}$O$(\alpha,\gamma)^{21}$Ne reaction rate was investigated using a one-zone post-processing nucleosynthesis code mimicking the core helium burning phase of a rotating massive star. The adopted temperature and density profiles follow the central shell of a $20M_{\odot}$ star with metalicity $z=10^{-5}$, computed with the Geneva stellar evolution code \cite{eggenberger2008geneva}. The initial abundances are extracted from the stellar core at the onset of core helium burning. During this one-zone calculation, $^{13}$C and $^{14}$N are injected at a constant rate to mimic the effect of rotation-induced mixing, a more detailed description of this method can be found in Ref. \cite{choplin2016}. 

\begin{figure}[ht!]
 \centering
 \includegraphics[width=0.48\textwidth]{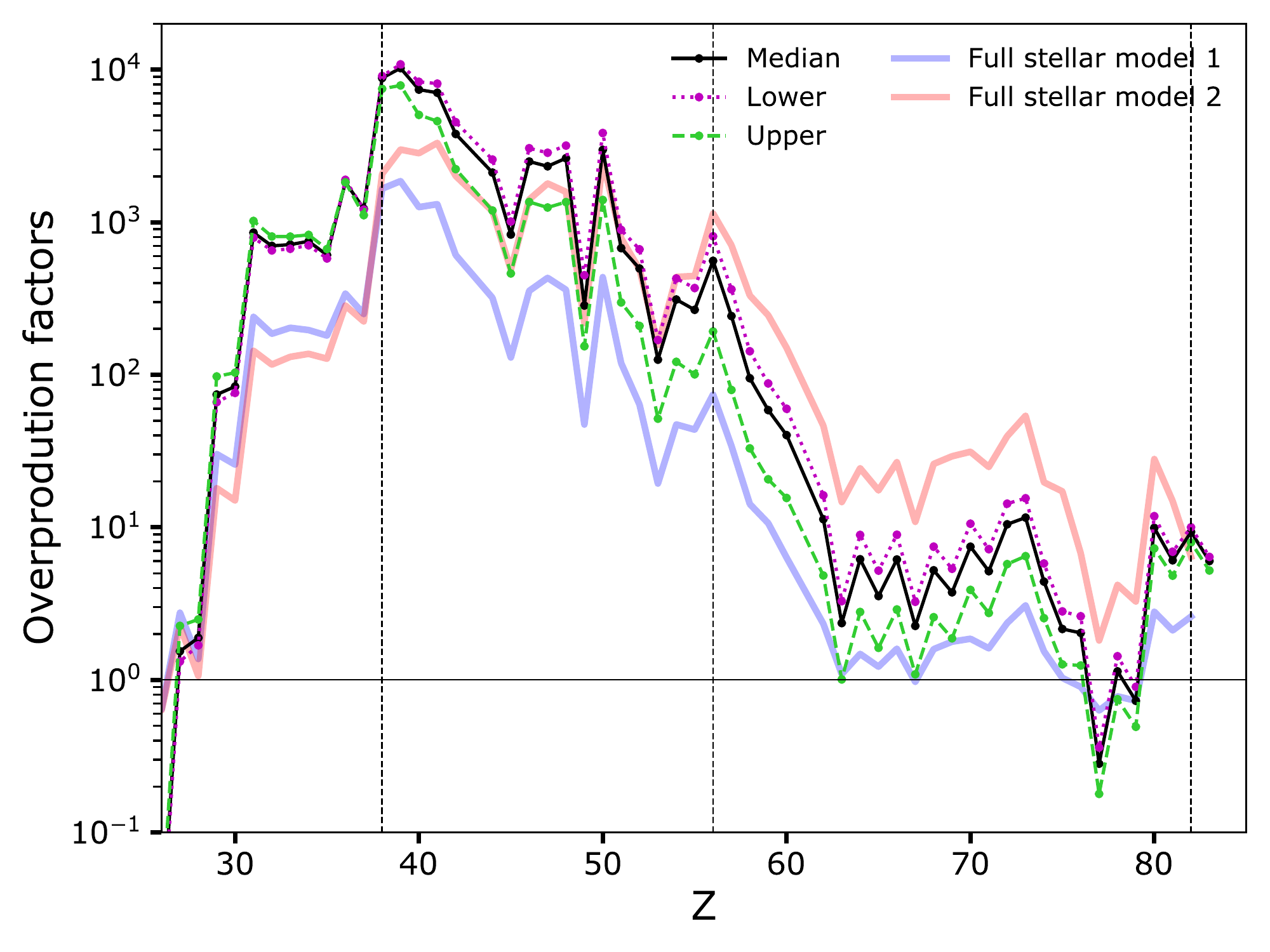}
\caption{s-process yields assuming different $^{17}$O$(\alpha,\gamma)^{21}$Ne and $^{17}$O$(\alpha,n)^{20}$Ne reaction rates. The lines denoted in the legend as lower, median and upper, represent the lower, median and upper $^{17}$O$(\alpha,\gamma)^{21}$Ne rates from the present work, assuming the recommended $^{17}$O$(\alpha,n)^{20}$Ne rate given by Best \textit{et al.} \cite{best2013}. The light blue and red patterns show yields from fast rotating 25 M$_{\odot}$ stellar models published in Ref \cite{choplin2018} (model 25S7 and 25S7B in their Table 1 and Figure 6). These stellar models were computed with either the recommended $^{17}$O$(\alpha,\gamma)^{21}$Ne rate from Best \textit{et al.} \cite{best2013} (blue) or this same rate but divided by ten (red). The four vertical lines show the location of Fe $(Z=26)$, Sr $(Z=38)$, Ba $(Z=56)$ and Pb $(Z=82)$.} \label{fig:sprocess_yeilds}
\end{figure}

Overproduction factors for elements in the range $25 < Z < 85$ are displayed on Figure \ref{fig:sprocess_yeilds}. The overproduction factor is defined as the final abundance divided by the initial abundance. Also shown in Figure \ref{fig:sprocess_yeilds} are the final s-process yields (integrated chemical composition ejected through winds and supernova) predicted from the two fast rotating 25 M$_{\odot}$ stellar models published in Ref \cite{choplin2018}. The present work suggests that those models, which assume the recommended $^{17}$O$(\alpha,\gamma)^{21}$Ne rate from Best \textit{et al.} (blue), and the same rate divided by 10 (red), represent robust lower and upper limits for s-process yields in terms of their dependence on the $^{17}$O$(\alpha,\gamma)^{21}$Ne rate uncertainties.

\section{Conclusion}

Key resonances in the $^{17}$O$(\alpha,\gamma)^{21}$Ne reaction were measured directly in inverse kinematics, including resonances at $E_{cm}=$ 1318, 814, 720, 638 and 613 keV. The strengths reported here for the 814, 721 and 638 keV resonances supersede those presented by Taggart \textit{et al.} \cite{taggart2019}. This is the first direct measurement of the 613 keV resonance, for which we place an upper limit that excludes any significant contribution to the overall reaction rate. The uncertainties for both the 638 and 721 keV resonance have been reduced from 78\% and 80\% to 16\% and 25\%, respectively. A new thermonuclear reaction rate was calculated, based on the resonance strengths presented here, which is approximately a factor of four to five lower than the previous rate at 0.3 GK \cite{best2013}. 

The impact of uncertainties in the present $^{17}$O$(\alpha,\gamma)^{21}$Ne rate were assessed using a post-processing code based on a $20M_{\odot}$ rotating star with low metallicity. The predicted yields lie between those assuming the $^{17}$O$(\alpha,\gamma)^{21}$Ne rate recommended by Best \textit{et al.} \cite{best2013} and the same rate divided by 10. The only remaining significant uncertainty in the $^{17}$O$(\alpha,\gamma)^{21}$Ne rate arises due to the as-yet unmeasured resonance at $E_{c.m.} = 305$ keV. Enhancement in s-process yields would be expected if the 305 keV resonance were found to have a width significantly below the estimated value. A direct measurement of this resonance would not be feasible with present beam intensities; however, indirect $\alpha$-transfer studies could present a viable way of addressing this one significant remaining uncertainty.

\section*{Acknowledgements}

We thank the excellent work of ISAC beam delivery, in particular OLIS personal led by K. Jayamanna, for providing the intense $^{17}$O beam that made this work possible. In addition, we acknowledge that this experiment was carried out during the COVID-19 pandemic and thank all TRIUMF staff in ensuring the experiment was carried out safely by implementing/observing measures to minimise risk of infection and transmission of the SARS-CoV-2 disease. TRIUMF’s core operations are supported via a contribution from the federal government through the National Research Council Canada, and the Government of British Columbia provides building capital funds. DRAGON is supported by funds from the Canadian Natural Science and Engineering Research Council (NSERC) under project number SAPPJ-2019-00039. UK authors gratefully acknowledge support provided by the Science and Technology Facilities Council (STFC). AC acknowledges support from the Fonds de la Recherche Scientifique-FNRS under Grant No IISN 4.4502.19. Authors from the Colorado School of Mines acknowledge funding via the U.S. Department of Energy Grant No. DE-FG02-93ER40789.

\bibliography{O17ag_references}

\appendix

\section{Thermonuclear Reaction Rate}
\label{sec:reactionrate}

This appendix contains the total thermonuclear reaction rate adopted in this work. The thermonuclear rate was computed using the \texttt{RatesMC} code, which calculates the log-normal parameters $\mu$ and $\sigma$ describing the reaction rate at a given temperature. Lower and upper rates are calculated at the 68\% confidence interval. The column labelled `A-D statistic' refers to the Anderson-Darling statistic, indicating how well a log-normal distribution describes the rate at a given temperature. An A-D statistic of less than $\approx1$ indicates that the rate is well described by a log-normal distribution. However, it has been shown that the assumption of a log-normal distributed reaction rate holds for A-D statistics in the $\approx 1-30$ range \cite{iliadis2015}.

{\setlength{\tabcolsep}{8pt}

\begin{longtable*}{c c c c c c c}

\caption{Tabulated thermonuclear reaction rate for the $^{17}$O$(\alpha,\gamma)^{21}$Ne reaction determined from the present work, expressed in units of cm$^{3}$ mol$^{-1}$ s$^{-1}$.} 

\label{table:reactionrate} \\

\toprule[1pt]\midrule[0.3pt]

 T [GK] & Low rate & Medium rate & High rate & Log-normal $\mu$ & Log-normal $\sigma$ & A-D statistic \\ \hline \noalign{\smallskip}
\endfirsthead

\multicolumn{7}{c}%
{{\tablename\ \thetable{} -- \textit{Continued from previous page}}} \\
\toprule[1pt]\midrule[0.3pt]
T [GK] & Low rate & Medium rate & High rate & Log-normal $\mu$ & Log-normal $\sigma$ & A-D statistic \\ \hline \noalign{\smallskip}
\endhead

\hline \noalign{\smallskip} \multicolumn{7}{c}{{\textit{Continued on next page}}} \\ \noalign{\smallskip} \hline
\endfoot

\midrule[0.3pt]\bottomrule[1pt]
\endlastfoot

0.010 & 1.29$\times$10$^{-67}$ & 1.69$\times$10$^{-66}$ &
      1.08$\times$10$^{-65}$ &  -1.518$\times$10$^{+02}$ &
       2.49$\times$10$^{+00}$  & 5.57$\times$10$^{+01}$ \\ 
0.011 & 2.93$\times$10$^{-64}$ & 3.67$\times$10$^{-63}$ &
      2.12$\times$10$^{-62}$ &  -1.442$\times$10$^{+02}$ &
       2.42$\times$10$^{+00}$  & 6.63$\times$10$^{+01}$ \\ 
0.012 & 1.80$\times$10$^{-61}$ & 2.18$\times$10$^{-60}$ &
      1.19$\times$10$^{-59}$ &  -1.378$\times$10$^{+02}$ &
       2.38$\times$10$^{+00}$  & 7.48$\times$10$^{+01}$ \\ 
0.013 & 4.06$\times$10$^{-59}$ & 4.77$\times$10$^{-58}$ &
      2.50$\times$10$^{-57}$ &  -1.324$\times$10$^{+02}$ &
       2.35$\times$10$^{+00}$  & 8.11$\times$10$^{+01}$ \\ 
0.014 & 4.11$\times$10$^{-57}$ & 4.81$\times$10$^{-56}$ &
      2.39$\times$10$^{-55}$ &  -1.278$\times$10$^{+02}$ &
       2.33$\times$10$^{+00}$  & 8.52$\times$10$^{+01}$ \\ 
0.015 & 2.24$\times$10$^{-55}$ & 2.61$\times$10$^{-54}$ &
      1.26$\times$10$^{-53}$ &  -1.238$\times$10$^{+02}$ &
       2.32$\times$10$^{+00}$  & 8.75$\times$10$^{+01}$ \\ 
0.016 & 7.42$\times$10$^{-54}$ & 8.37$\times$10$^{-53}$ &
      4.06$\times$10$^{-52}$ &  -1.203$\times$10$^{+02}$ &
       2.31$\times$10$^{+00}$  & 8.80$\times$10$^{+01}$ \\ 
0.018 & 2.40$\times$10$^{-51}$ & 2.72$\times$10$^{-50}$ &
      1.36$\times$10$^{-49}$ &  -1.146$\times$10$^{+02}$ &
       2.29$\times$10$^{+00}$  & 8.42$\times$10$^{+01}$ \\ 
0.020 & 2.45$\times$10$^{-49}$ & 2.68$\times$10$^{-48}$ &
      1.38$\times$10$^{-47}$ &  -1.099$\times$10$^{+02}$ &
       2.22$\times$10$^{+00}$  & 7.03$\times$10$^{+01}$ \\ 
0.025 & 1.79$\times$10$^{-45}$ & 1.11$\times$10$^{-44}$ &
      5.70$\times$10$^{-44}$ &  -1.013$\times$10$^{+02}$ &
       1.74$\times$10$^{+00}$  & 1.09$\times$10$^{+01}$ \\ 
0.030 & 2.29$\times$10$^{-42}$ & 9.77$\times$10$^{-42}$ &
      3.31$\times$10$^{-41}$ &  -9.456$\times$10$^{+01}$ &
       1.43$\times$10$^{+00}$  & 2.73$\times$10$^{+01}$ \\ 
0.040 & 5.34$\times$10$^{-38}$ & 3.66$\times$10$^{-37}$ &
      1.65$\times$10$^{-36}$ &  -8.407$\times$10$^{+01}$ &
       1.68$\times$10$^{+00}$  & 1.54$\times$10$^{+01}$ \\ 
0.050 & 1.09$\times$10$^{-33}$ & 3.43$\times$10$^{-33}$ &
      9.18$\times$10$^{-33}$ &  -7.484$\times$10$^{+01}$ &
       1.10$\times$10$^{+00}$  & 1.11$\times$10$^{+01}$ \\ 
0.060 & 4.72$\times$10$^{-30}$ & 1.79$\times$10$^{-29}$ &
      6.02$\times$10$^{-29}$ &  -6.626$\times$10$^{+01}$ &
       1.30$\times$10$^{+00}$  & 5.47$\times$10$^{+00}$ \\ 
0.070 & 4.20$\times$10$^{-27}$ & 1.50$\times$10$^{-26}$ &
      4.94$\times$10$^{-26}$ &  -5.952$\times$10$^{+01}$ &
       1.27$\times$10$^{+00}$  & 5.85$\times$10$^{+00}$ \\ 
0.080 & 8.66$\times$10$^{-25}$ & 3.26$\times$10$^{-24}$ &
      9.34$\times$10$^{-24}$ &  -5.422$\times$10$^{+01}$ &
       1.25$\times$10$^{+00}$  & 2.40$\times$10$^{+01}$ \\ 
0.090 & 6.02$\times$10$^{-23}$ & 2.39$\times$10$^{-22}$ &
      6.52$\times$10$^{-22}$ &  -4.997$\times$10$^{+01}$ &
       1.27$\times$10$^{+00}$  & 3.90$\times$10$^{+01}$ \\ 
0.100 & 1.84$\times$10$^{-21}$ & 7.69$\times$10$^{-21}$ &
      2.19$\times$10$^{-20}$ &  -4.651$\times$10$^{+01}$ &
       1.32$\times$10$^{+00}$  & 4.06$\times$10$^{+01}$ \\ 
0.110 & 3.03$\times$10$^{-20}$ & 1.34$\times$10$^{-19}$ &
      4.11$\times$10$^{-19}$ &  -4.364$\times$10$^{+01}$ &
       1.38$\times$10$^{+00}$  & 3.69$\times$10$^{+01}$ \\ 
0.120 & 3.09$\times$10$^{-19}$ & 1.46$\times$10$^{-18}$ &
      4.87$\times$10$^{-18}$ &  -4.124$\times$10$^{+01}$ &
       1.44$\times$10$^{+00}$  & 3.33$\times$10$^{+01}$ \\ 
0.130 & 2.23$\times$10$^{-18}$ & 1.15$\times$10$^{-17}$ &
      4.02$\times$10$^{-17}$ &  -3.919$\times$10$^{+01}$ &
       1.50$\times$10$^{+00}$  & 3.13$\times$10$^{+01}$ \\ 
0.140 & 1.18$\times$10$^{-17}$ & 6.67$\times$10$^{-17}$ &
      2.45$\times$10$^{-16}$ &  -3.744$\times$10$^{+01}$ &
       1.54$\times$10$^{+00}$  & 3.08$\times$10$^{+01}$ \\ 
0.150 & 5.16$\times$10$^{-17}$ & 3.12$\times$10$^{-16}$ &
      1.17$\times$10$^{-15}$ &  -3.591$\times$10$^{+01}$ &
       1.58$\times$10$^{+00}$  & 3.11$\times$10$^{+01}$ \\ 
0.160 & 1.86$\times$10$^{-16}$ & 1.19$\times$10$^{-15}$ &
      4.60$\times$10$^{-15}$ &  -3.457$\times$10$^{+01}$ &
       1.60$\times$10$^{+00}$  & 3.17$\times$10$^{+01}$ \\ 
0.180 & 1.61$\times$10$^{-15}$ & 1.13$\times$10$^{-14}$ &
      4.47$\times$10$^{-14}$ &  -3.234$\times$10$^{+01}$ &
       1.62$\times$10$^{+00}$  & 3.24$\times$10$^{+01}$ \\ 
0.200 & 9.54$\times$10$^{-15}$ & 6.78$\times$10$^{-14}$ &
      2.73$\times$10$^{-13}$ &  -3.053$\times$10$^{+01}$ &
       1.60$\times$10$^{+00}$  & 3.06$\times$10$^{+01}$ \\ 
0.250 & 5.11$\times$10$^{-13}$ & 1.96$\times$10$^{-12}$ &
      6.97$\times$10$^{-12}$ &  -2.695$\times$10$^{+01}$ &
       1.18$\times$10$^{+00}$  & 2.86$\times$10$^{+01}$ \\ 
0.300 & 4.67$\times$10$^{-11}$ & 5.96$\times$10$^{-11}$ &
      1.01$\times$10$^{-10}$ &  -2.342$\times$10$^{+01}$ &
       4.09$\times$10$^{-01}$  & 1.53$\times$10$^{+02}$ \\ 
0.350 & 2.29$\times$10$^{-09}$ & 2.48$\times$10$^{-09}$ &
      2.73$\times$10$^{-09}$ &  -1.981$\times$10$^{+01}$ &
       9.56$\times$10$^{-02}$  & 1.77$\times$10$^{+01}$ \\ 
0.400 & 4.76$\times$10$^{-08}$ & 5.12$\times$10$^{-08}$ &
      5.50$\times$10$^{-08}$ &  -1.679$\times$10$^{+01}$ &
       7.17$\times$10$^{-02}$  & 2.20$\times$10$^{-01}$ \\ 
0.450 & 5.22$\times$10$^{-07}$ & 5.60$\times$10$^{-07}$ &
      6.00$\times$10$^{-07}$ &  -1.440$\times$10$^{+01}$ &
       7.05$\times$10$^{-02}$  & 1.40$\times$10$^{-01}$ \\ 
0.500 & 3.54$\times$10$^{-06}$ & 3.80$\times$10$^{-06}$ &
      4.07$\times$10$^{-06}$ &  -1.248$\times$10$^{+01}$ &
       6.96$\times$10$^{-02}$  & 1.53$\times$10$^{-01}$ \\ 
0.600 & 6.15$\times$10$^{-05}$ & 6.58$\times$10$^{-05}$ &
      7.02$\times$10$^{-05}$ &  -9.629$\times$10$^{+00}$ &
       6.77$\times$10$^{-02}$  & 2.55$\times$10$^{-01}$ \\ 
0.700 & 4.60$\times$10$^{-04}$ & 4.92$\times$10$^{-04}$ &
      5.24$\times$10$^{-04}$ &  -7.618$\times$10$^{+00}$ &
       6.60$\times$10$^{-02}$  & 3.63$\times$10$^{-01}$ \\ 
0.800 & 2.04$\times$10$^{-03}$ & 2.18$\times$10$^{-03}$ &
      2.32$\times$10$^{-03}$ &  -6.129$\times$10$^{+00}$ &
       6.43$\times$10$^{-02}$  & 4.30$\times$10$^{-01}$ \\ 
0.900 & 6.44$\times$10$^{-03}$ & 6.87$\times$10$^{-03}$ &
      7.30$\times$10$^{-03}$ &  -4.982$\times$10$^{+00}$ &
       6.26$\times$10$^{-02}$  & 4.16$\times$10$^{-01}$ \\ 
1.000 & 1.61$\times$10$^{-02}$ & 1.71$\times$10$^{-02}$ &
      1.82$\times$10$^{-02}$ &  -4.068$\times$10$^{+00}$ &
       6.06$\times$10$^{-02}$  & 3.70$\times$10$^{-01}$ \\ 
1.250 & 8.50$\times$10$^{-02}$ & 8.97$\times$10$^{-02}$ &
      9.46$\times$10$^{-02}$ &  -2.411$\times$10$^{+00}$ &
       5.45$\times$10$^{-02}$  & 3.86$\times$10$^{-01}$ \\ 
1.500 & 2.67$\times$10$^{-01}$ & 2.81$\times$10$^{-01}$ &
      2.95$\times$10$^{-01}$ &  -1.271$\times$10$^{+00}$ &
       4.92$\times$10$^{-02}$  & 3.73$\times$10$^{-01}$ \\ 
1.750 & 6.25$\times$10$^{-01}$ & 6.55$\times$10$^{-01}$ &
      6.86$\times$10$^{-01}$ &  -4.232$\times$10$^{-01}$ &
       4.68$\times$10$^{-02}$  & 3.07$\times$10$^{-01}$ \\ 
2.000 & 1.20$\times$10$^{+00}$ & 1.26$\times$10$^{+00}$ &
      1.32$\times$10$^{+00}$ &  2.326$\times$10$^{-01}$ &
       4.70$\times$10$^{-02}$  & 3.64$\times$10$^{-01}$ \\ 
2.500 & 3.06$\times$10$^{+00}$ & 3.22$\times$10$^{+00}$ &
      3.38$\times$10$^{+00}$ &  1.168$\times$10$^{+00}$ &
       5.05$\times$10$^{-02}$  & 4.34$\times$10$^{-01}$ \\ 
3.000 & 5.65$\times$10$^{+00}$ & 5.96$\times$10$^{+00}$ &
      6.30$\times$10$^{+00}$ &  1.786$\times$10$^{+00}$ &
       5.42$\times$10$^{-02}$  & 5.34$\times$10$^{-01}$ \\ 
3.500 & 8.62$\times$10$^{+00}$ & 9.11$\times$10$^{+00}$ &
      9.65$\times$10$^{+00}$ &  2.210$\times$10$^{+00}$ &
       5.70$\times$10$^{-02}$  & 5.50$\times$10$^{-01}$ \\ 
4.000 & 1.16$\times$10$^{+01}$ & 1.23$\times$10$^{+01}$ &
      1.31$\times$10$^{+01}$ &  2.511$\times$10$^{+00}$ &
       5.90$\times$10$^{-02}$  & 5.41$\times$10$^{-01}$ \\ 
5.000 & 1.69$\times$10$^{+01}$ & 1.79$\times$10$^{+01}$ &
      1.91$\times$10$^{+01}$ &  2.888$\times$10$^{+00}$ &
       6.15$\times$10$^{-02}$  & 5.48$\times$10$^{-01}$ \\ 
6.000 & 2.08$\times$10$^{+01}$ & 2.21$\times$10$^{+01}$ &
      2.35$\times$10$^{+01}$ &  3.096$\times$10$^{+00}$ &
       6.31$\times$10$^{-02}$  & 5.50$\times$10$^{-01}$ \\ 
7.000 & 2.33$\times$10$^{+01}$ & 2.48$\times$10$^{+01}$ &
      2.64$\times$10$^{+01}$ &  3.212$\times$10$^{+00}$ &
       6.41$\times$10$^{-02}$  & 5.45$\times$10$^{-01}$ \\ 
8.000 & 2.48$\times$10$^{+01}$ & 2.63$\times$10$^{+01}$ &
      2.81$\times$10$^{+01}$ &  3.272$\times$10$^{+00}$ &
       6.47$\times$10$^{-02}$  & 5.42$\times$10$^{-01}$ \\ 
9.000 & 2.54$\times$10$^{+01}$ & 2.71$\times$10$^{+01}$ &
      2.89$\times$10$^{+01}$ &  3.300$\times$10$^{+00}$ &
       6.52$\times$10$^{-02}$  & 5.39$\times$10$^{-01}$ \\ 
10.000 & 2.56$\times$10$^{+01}$ & 2.72$\times$10$^{+01}$ &
      2.91$\times$10$^{+01}$ &  3.305$\times$10$^{+00}$ &
       6.56$\times$10$^{-02}$  & 5.35$\times$10$^{-01}$ \\

\end{longtable*}
}

\end{document}